\documentclass[floatfix,twocolumn,epsf,prb,amsmath,amssymb,showpacs,superscriptaddress]{revtex4}
\usepackage{graphicx,color,dsfont}
\usepackage{subfigure}
\begin{document}
\newcommand{\ud}{\mathrm{d}}
\newcommand{\kvec}[2]{\begin{pmatrix} #1 \\ #2 \end{pmatrix} }
\newcommand{\kvecc}[4]{\begin{pmatrix} #1 \\ #2 \\ #3 \\ #4 \end{pmatrix} }
\newcommand{\kvecje}[2]{\bigl( \begin{smallmatrix} #1 \\ #2 \end{smallmatrix} \bigr)}
\newcommand{\rvec}[2]{\begin{pmatrix}#1 & #2 \end{pmatrix} }
\newcommand{\rvecje}[2]{(\begin{smallmatrix} #1 & #2 \end{smallmatrix})}
\newcommand{\matt}[4]{\begin{pmatrix} #1 & #2 \\ #3 & #4 \end{pmatrix} }
\newcommand{\mattje}[4]{\bigl( \begin{smallmatrix} #1 & #2 \\ #3 & #4 \end{smallmatrix} \bigr)}
\newcommand{\vgll}[2]{\left\{ \begin{array}{c} #1 \\ #2 \end{array} \right. }
\newcommand{\vglL}[2]{\begin{equation} \left\{ \begin{align} #1 \\ #2 \end{align} \end{equation} \right. }
\newcommand{\pd}[2]{\frac{\partial #1}{\partial #2}}
\newcommand{\pdd}[2]{\frac{{\partial}^2 #1}{{\partial #2}^2}}
\newcommand{\pdje}[1]{\partial_{#1}}
\newcommand{\ve}{\varepsilon}
\title{Bilayer graphene with single and multiple electrostatic barriers:\\
 band structure and transmission}
\author{Micha\"el Barbier}
\email{michael.barbier@gmail.com}
\affiliation{Department of Physics, University of Antwerp, Groenenborgerlaan 171, B-2020 Antwerpen, Belgium\\}
\author{P.~Vasilopoulos}
\email{takis@alcor.concordia.ca}
\affiliation{Department of Physics, Concordia University, 7141 Sherbrooke Ouest, Montr\'eal, Quebec, Canada H4B 1R6\\}
\author{F.~M.~Peeters}
\email{francois.peeters@ua.ac.be}
\affiliation{Department of Physics, University of Antwerp, Groenenborgerlaan 171, B-2020 Antwerpen, Belgium\\}
\author{J. Milton Pereira Jr.}
\email{joaomilton.pereira@ua.ac.be}
\affiliation{Departamento de F\'{\i}sica, Universidade
Federal do Cear\'a, Fortaleza, Cear\'a $60455$-$760$, Brazil\\}
\begin{abstract}
We evaluate the electronic transmission and conductance in bilayer graphene
 through a finite number of potential barriers.  Further, we evaluate
the dispersion relation in a bilayer graphene superlattice with a
periodic potential applied to both layers. As model we use the massless Dirac-Weyl equation in the continuum model. For zero bias the
dispersion relation shows a finite gap for carriers with zero
momentum in the direction parallel to the barriers. This is in contrast to
single-layer graphene where no such gap was found. A gap also appears for a
finite bias.
Numerical results for the energy spectrum, conductance, and the
density of states are presented and contrasted with those pertaining to
single-layer graphene.
\end{abstract}
 \maketitle
\section{Introduction}
Low-dimensional systems have long been the subject of intensive
research, both on their fundamental properties and on
possible applications. In this respect  the recent production of
atom-thick crystal carbon layers (graphene) has raised the
possibility of the development of new graphene-based devices that
exploit its unusual electronic and mechanical properties (for a
recent review  see Ref.~\onlinecite{cast}).
The electronic spectrum of defect-free single-layer graphene is
gapless and, together with the chiral aspect of the carriers in this
system, leads to a perfect transmission through 
an arbitrarily high and wide 
potential barrier, i.e.,  the Klein paradox\cite{klein,kats}. That 
can be avoided if a gap is introduced in the electronic
spectrum 
and may be necessary for certain
applications, e. g., for improving the on/off ratio in
carbon-based transistors.

There are a few methods to introduce a gap in the spectrum of graphene. One
of them is to use nanoribbons
in  which a bandgap \cite{loss} arises  due to the lateral
confinement.
Also, depositing graphene on a substrate such as boron nitride
was found recently to result in a bandgap\cite{boron} of $53$ meV. In bilayer
graphene\cite{partoens} a gap can be introduced by applying a bias
between the two layers or by doping one of them such that  a
potential difference results between the layers
\cite{ohta,castro,mccann2,mccann}. Changing the bias in the latter
case can open and close the gap dynamically which is interesting
for transistor applications. Nanostructured gates
can thus allow the creation of quantum dots on bilayer graphene
\cite{milton2}.

In this paper we investigate the electronic properties of a biased
bilayer in which the potential difference between the two layers is changed periodically. Such a superlattice (SL), which can
be created by applying gates to the bilayer, 
is of interest as it 
shows how a one-dimensional  band structure may appear in such
a system. An additional motivation is that curvature effects of
corrugated single-layer graphene lead to an effective periodic
potential resembling that of a SL\cite{corrugated}. Although in a
bilayer this effect would be 
weaker, since the bilayer is less bendable than a single layer,
it might  still be important.

The paper is organized as follows. Section II briefly shows the
basic formalism. In Sec.~III results for the transmission and
conductance through a finite number of barriers are presented.
Section IV shows results for the dispersion relation and the
density of states in SLs in bilayer graphene. Finally, a
summary and concluding remarks are 
given in Sec.~V.
\section{Hamiltonian, energy spectrum, and eigenstates}
Bilayer graphene consists of two A-B-stacked monolayers of graphene. 
Each monolayer has two independent atoms A and B in its
unit cell. The relevant Hamiltonian, obtained by a
nearest-neighbour, tight-binding approximation near the K-point and
the eigenstates $\Psi$ read
\begin{equation}\label{eq_1}
	\mathcal{H} = 
		\begin{pmatrix}
			V_1 & v_F\pi & t_\perp & 0 \\
			v_F\pi^\dagger & V_1 & 0 & 0 \\
			t_\perp & 0 & V_2 & v_F\pi^\dagger \\
			0 & 0 & v_F\pi & V_2 
		\end{pmatrix}, 
	\quad \quad \psi = 
		\begin{pmatrix} 
			\psi_A \\ \psi_B \\ \psi_{B'} \\ \psi_{A'}
	  \end{pmatrix}.
\end{equation}
Here  $\pi = (p_x + i p_y)$, $p_{x,y} = - i \hbar \partial_{x,y}$ is
the momentum operator, $v_F = 10^6$ m/s is the Fermi velocity, $V_1$ and $V_2$ are the potentials on layers
$1$ and $2$, respectively, and $t_\perp$ describes the coupling between
these layers. As shown in  Appendix A, for   spatially independent $t_\perp$, $V_1$, and
$V_2$, the spectrum consists of four bands given by
\begin{equation}
\begin{aligned}
	\ve_\pm^{'+} &= \left[  \epsilon_{kt'}^2 
	\pm t' \sqrt{ 4k^2\delta^2/t'^2 + k^2 + t'^2/4 } \right]^{1/2},\\
	\ve_\pm^{'-} &= - \left[  \epsilon_{kt'}^2 
	\pm t' \sqrt{ 4k^2\delta^2/t'^2 + k^2 + t'^2/4 } \right]^{1/2}.
\end{aligned}
\end{equation}
Here $ \epsilon_{kt'}^2=k^2 + \delta^2 + t'^2/2$,  $\Delta=(V_1 - V_2)$, $\delta = \Delta/2 \hbar v_F$,  $\ve = E / \hbar v_F$ and $t' = t_\perp
/ \hbar v_F$. The eigenstates $ \psi $ of $\mathcal{H}$
 are given by Eq. (\ref{app_ev}) in the Appendix.

A reduced version of the 
four-band Hamiltonian shown in Eq. (1) that
is often used\cite{mccann2} is given by
\begin{equation}\label{eq_2}
	\mathcal{H} = 
	- \frac{{v_F}^2}{t_\perp}
		\begin{pmatrix}
			V & {\pi^\dagger}^2 \\
			\pi^2 & V
		\end{pmatrix},
\end{equation}
Assuming solutions of the form $A \exp(i k_x x)\exp(i k_y y)$ we can replace $p_x$ by $\hbar k_x$. Then setting the determinant of the equation
$\mathcal{H} \psi = E \psi$ equal to zero gives 
rise to the two-band spectrum
\begin{equation}
    E-V = \pm ({v_F}^2 \hbar^2/t_\perp)(k_x^2 + k_y^2),
\end{equation}
where $V$ is the potential applied to each layer. In the next
section we 
compare some of the results obtained from Eqs.~(1)
and (2) with those obtained from Eqs.~(3) and (4).

\section{Finite number of barriers}
\subsection{Transmission}
%
\begin{figure}
     \begin{center}
       \includegraphics[height=4cm]{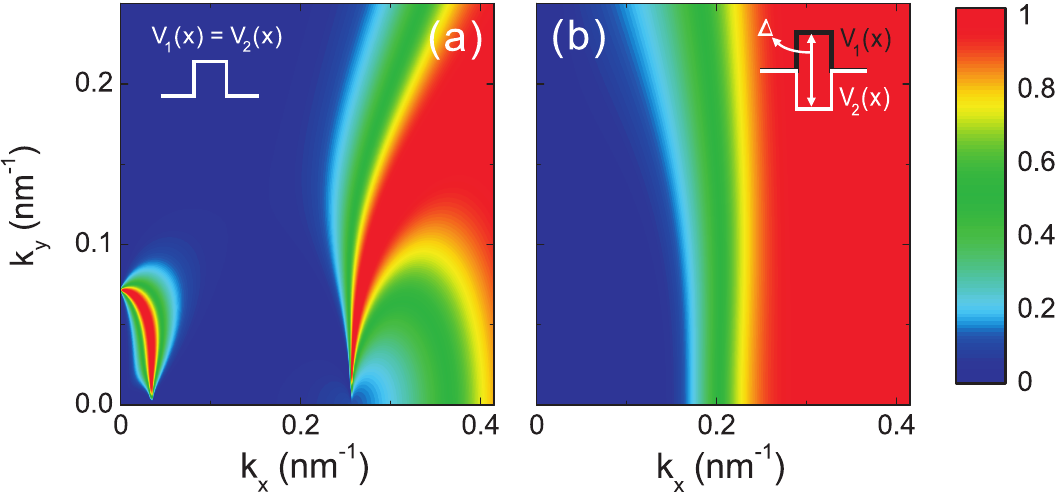}
    \end{center}
    \caption{(Color online) Transmission through a single square barrier of
    height $V = 100$ meV and width $10$ nm.
    The potential difference  $ \delta = (V_1-V_2)/2\hbar v_F$ 
    is zero
    in panel (a) and $\Delta = 100$ meV in panel (b), inside the barrier/well region.}
    \label{fig_trans1}
\end{figure}
\begin{figure}
     \begin{center}
    \includegraphics[height=4cm]{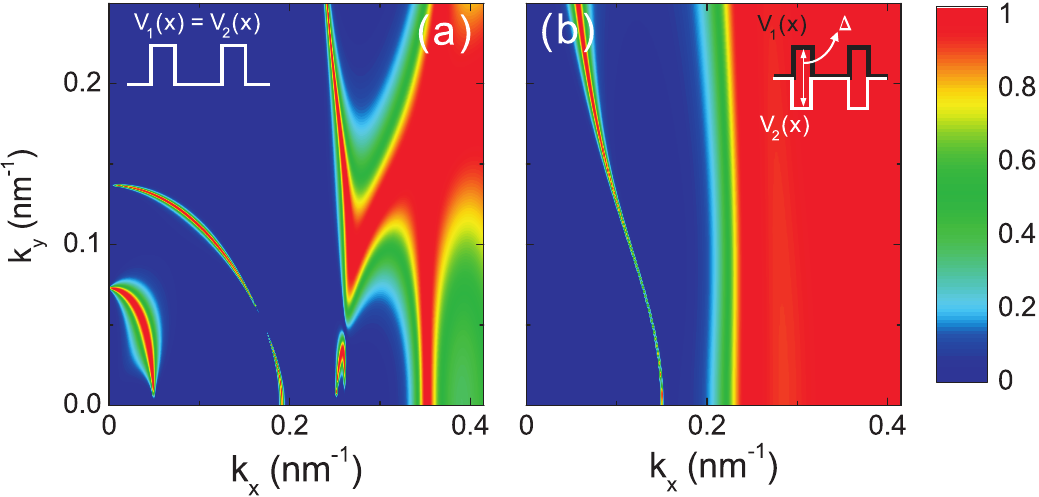}
    \end{center}
    \caption{(Color online) Transmission through a double barrier.
    The square barriers are $100$ meV high and  $10$ nm wide,  the  distance between
    them is $L = 10$ nm. In panel (a) the potential difference between the layers is zero,
    in panel (b) it is $100$ meV inside the barrier
    /well regions.}\label{fig_trans2}
\end{figure}
When applying a bias to a metallic strip a potential barrier is 
created; 
we will approximate it by a square potential barrier. The
eigenstates $\Psi$  given in Appendix A can be used in each region
of 
constant potential. In  matrix notation the wave function in
region j with a constant potential can be written as a matrix
product (cf.\ Eq.~(\ref{app_wavefct}) of the Appendix),
\begin{equation}
    \Psi_j = \mathcal{G}_j \mathcal{M}_j \mathcal{A}_j,
\end{equation}
 where
$\mathcal{A}_j = [A_j$, $B_j$, $C_j$, $D_j]^T$ and the superscript
$T$ denotes the transpose of the row vector. Then we apply  the
continuity of the wave function at the different potential steps.
For the $(j+1)^{th}$ potential step at $x_{j+1}$ we obtain
\begin{equation}
    \mathcal{A}_{j+1} = \mathcal{M}^{-1}_{j+1}(x_{j+1}) \mathcal{G}^{-1}_{j+1}
    \mathcal{G}_{j} \mathcal{M}_{j}(x_{j+1}) \mathcal{A}_{j}.
\end{equation}
This links the coefficients of the wave function behind the barriers
to those in front of them. Then 
we can write
\begin{equation}
    \mathcal{A}_{n+1} = \mathcal{N} \mathcal{A}_{1},
\end{equation}
where $\mathcal{N}_j = \mathcal{M}^{-1}_{j+1}(x_{j+1}) \mathcal{G}^{-1}_{j+1}
\mathcal{G}_{j} \mathcal{M}_{j}(x_{j+1})$ and $\mathcal{N} = \prod_j{\mathcal{N}_j}$.
From now on we assume   $|E| < t_\perp$ outside the barrier such that
$\alpha_+ \in \mathbb{R}$ and $\alpha_- \in \mathbb{C}$, see
the Appendix. Assuming that there is an incident wave, with wave
vector $\alpha_+$ from the left (normalized to unity),  part of it
will be reflected (coefficient r) and part of it will be transmitted
(coefficient t). Also there are growing and decaying evanescent
 states near the barrier (coefficients $e_g$ and
$e_d$, respectively). The relation between all these waves is
written in  the form
\begin{equation}
    \kvecc{t}{0}{e_d}{0} = \mathcal{N} \kvecc{1}{r}{0}{e_g};
\end{equation}
it can be rewritten as a linear system of equations,
\begin{equation}
    -\kvecc{N_{11}}{N_{21}}{N_{31}}{N_{41}} =
    \begin{pmatrix}
			-1 & N_{12} &  0 & N_{14} \\
			0 & N_{22} &  0 & N_{24} \\
			0 & N_{32} & -1 & N_{34}\\
			0 & N_{42} &  0 & N_{44}
    \end{pmatrix} \kvecc{t}{r}{e_d}{e_g},
\end{equation}
where $N_{ij}$ are the coefficient of $\mathcal{N}$. We solved this set of
equations numerically. The transmission is now given by $T = |t|^2$.

A contour plot of the transmission through a single barrier is shown
in Fig.~\ref{fig_trans1}. 
 Panel (a)
  is for a barrier with height $100$ meV and width $10$ nm and the potential
  difference $\Delta$ between the layers is zero. 
In contrast with the case of a 2DEG, there are transmission resonances 
in the region which corresponds to energies lower than the barrier height. These are due to the hole states inside the barrier through which the electrons can tunnel.
In panel (b) the barrier is 
$10$ nm wide and 
the potential difference between the layers is $\Delta = 100$ meV.
As can be seen, the $k_y$ dependence of the transmission in panel
(b) is weaker than that on $k_x$ and resembles more the
Schr\"odinger case. 
The transmission of 
electrons, at normal  incidence ($k_y=0$),
starts from a $k_x \simeq 0.23$ nm$^{-}$ meV which corresponds to an energy of
$50$ meV which is the edge of the gap,
 inside which
there  are evanescent states which supress the transmission. For the double barrier system we see that there is also a resonance at $k_x \simeq 0.16$ nm$^{-}$. In contrast with the single-layer case, in Fig. 1 (a) and 2 (a) there is no
perfect transmission for normal incidence ($k_y=0$), even though the system is
gapless. This is a consequence of the chiral nature of the carriers in bilayer
graphene (see, e.g. Ref. \onlinecite{kats}).

A plot of the transmission through a double
barrier is shown in Fig.~\ref{fig_trans2}. The 
barriers are $100$ meV high,   $10$ nm wide, and the  distance
between them is $L = 10$ nm. Panel (a) is for $\delta=0$ and panel
(b) for $\Delta = 100$ meV. In agreement with Ref.
\onlinecite{trans} we find that it is the distance L between the
barriers and
not their 
width that is important in determining the tunneling
states and thus the transmission. For more results, e.~g.\
conduction through {\it unbiased}, multiple-barrier systems
see Refs.~\onlinecite{trans} and ~\onlinecite{trans1}.

\begin{figure}
     \begin{center}
        \includegraphics[height=6cm]{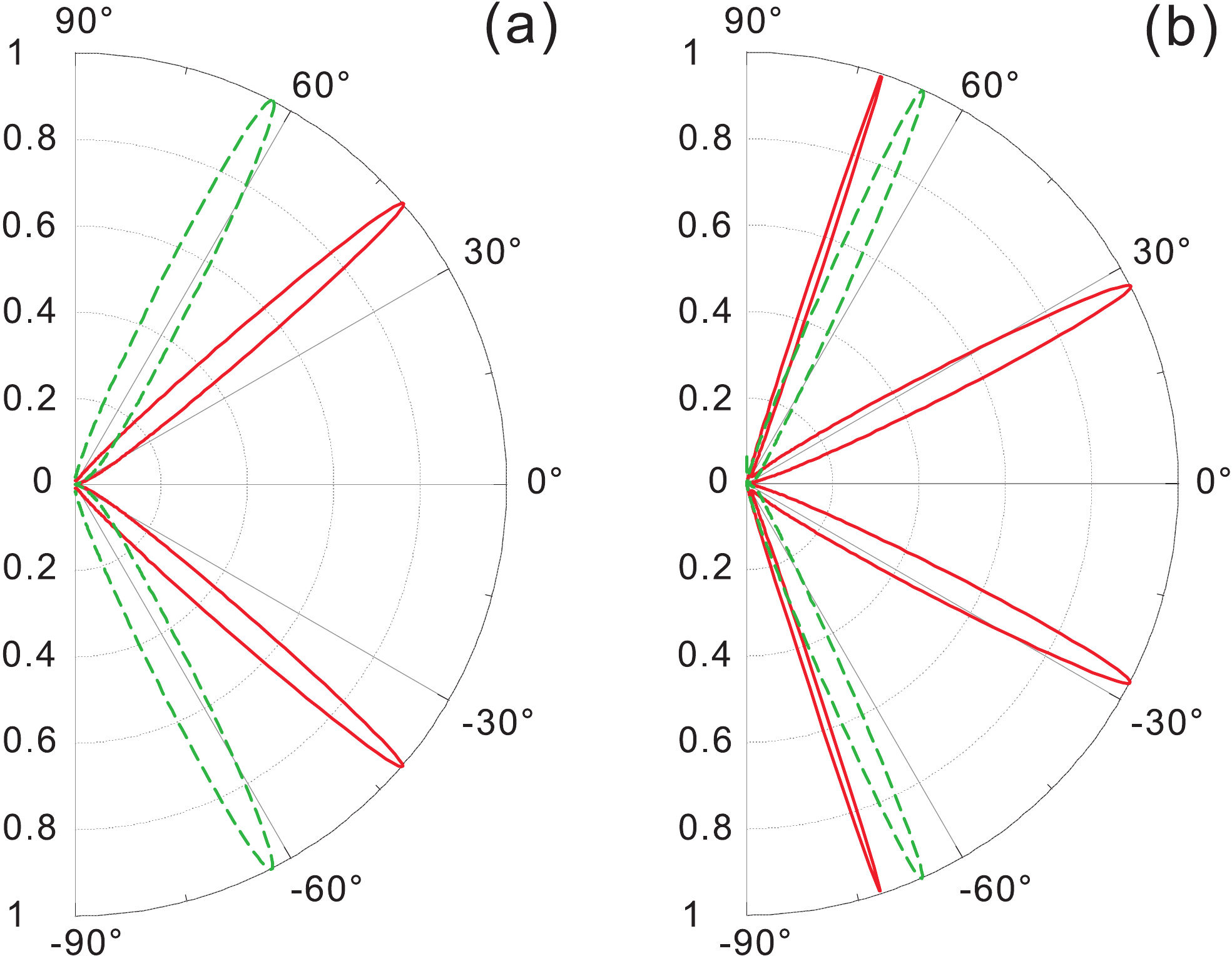}
     \end{center}
\vspace*{-0.2cm}
    \caption{(Color online) Transmission through a $100$ nm wide barrier as a function of the angle of
incidence for constant energy $E=17$ meV. Panel (a) 
 results from the
$4\times 4$ Hamiltonian of Eq.~(1) and panel (b) from the $2\times
2$ one of Eq.~(3). The solid red and dashed green curves are for a
single barrier with height $50$ meV and $100$ meV,
respectively.}\label{fig_trans3}
\end{figure}

The transmission shown in Fig.~\ref{fig_trans1} and \ref{fig_trans2} depends on
the angle of incidence $\phi$
given by $\tan \phi=k_y/k_x$. A more direct way to see this is shown in Fig.~\ref{fig_trans3}
where the transmission is plotted as a function of the angle of
incidence for constant energy $E=17$ meV. For panel (a) we used the
$4\times 4$ Hamiltonian of Eq.~(1), where a bias $\Delta = 0$, while for
panel (b) we used the $2\times 2$ one of Eq.~(3). The differences are due to the
different spectra of Eqs.~(2) and (4).
We consider the results of panel (a) to be more accurate than those of panel
(b), since the Hamiltonian of Eq. (1) includes the effect of the
 non-parabolicity
of the electron dispersion, which can have a strong effect on the resonant
transmission of the carriers.

\subsection{Conductance}
    It is  interesting to see to what extent the transmission
affects 
the conductance $G$, which is given by
\begin{equation}
  G = G_0 \int_{-\pi/2}^{\pi/2} T(E,\phi) \cos\phi d \phi.
\end{equation}
Here $G_{0} = 2 e^{2} \sqrt{E_{F}^2 + t_\perp E_{F}} L_y / (\pi h \hbar v_F)$, $\phi$ is the angle of
incidence measured from the $x$ axis, $T(E,\phi)$ the
transmission through the structure at energy $E$, and $L_y$  the
length of the structure along the $y$ direction.

\begin{figure}
     \begin{center}
\vspace*{-0.6cm}
        \includegraphics[height=6cm]{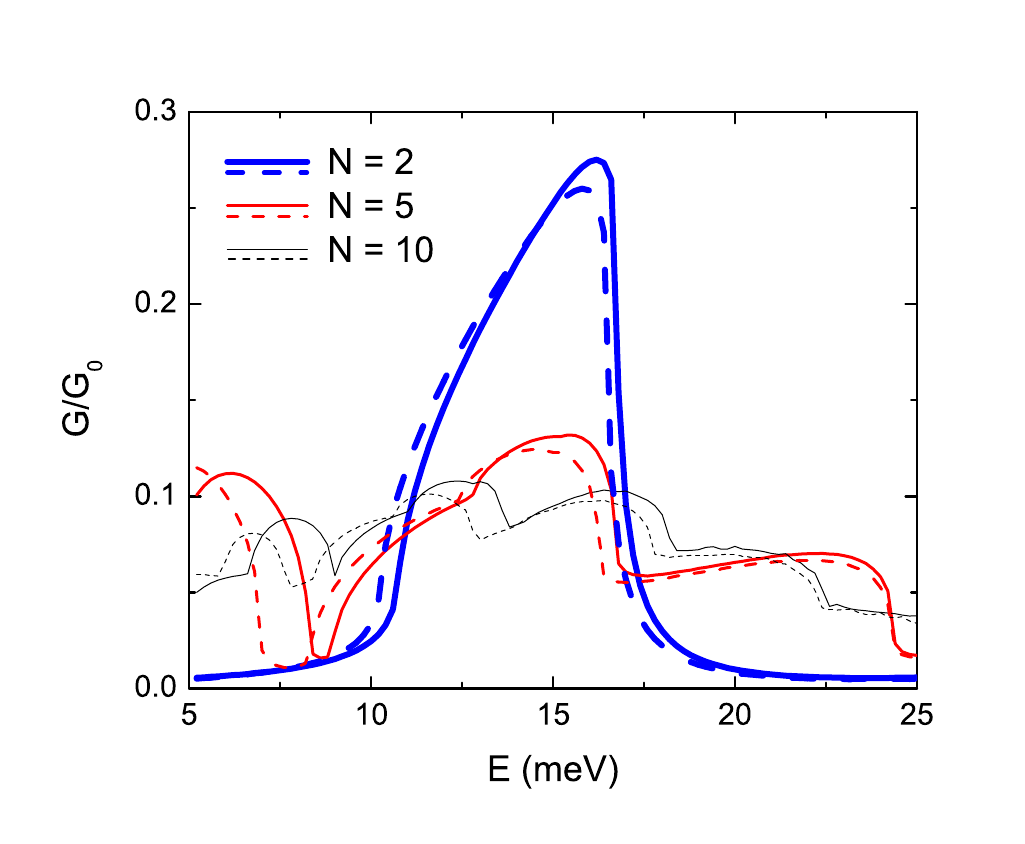}
    \end{center}
\vspace*{-0.85cm}
    \caption{(Color online) Conductance as a function of  energy. The very thick blue, thick red, and
    thin black solid curves are for two, five, and ten barriers, respectively, of width $10$ nm, height $100$ meV and with an interbarrier distance of $5$ nm, and result from the $4\times 4$ Hamiltonian, Eq. (\ref{eq_1}), while the dashed curves result from the $2\times 2$ one, Eq. (\ref{eq_2}).}\label{fig_cond1}
\end{figure}

In Fig. ~\ref{fig_cond1} we plot
 the conductance $G$ through two, five, and ten barriers in
blue, red, and black color, respectively. The height of the
barriers is $100$ meV, their width $D = 10$ nm, and the
interbarrier distance $L=5$ nm.
The solid curves are obtained  using Eq.~(1) and the dashed 
ones using the reduced Hamiltonian of Eq.~(3) with the
 same 
 coupling strength $t_\perp = 390$ meV. As can be
seen, both models give qualitatively the same results. The disagreement
is mostly apparent in the low-energy region and is mainly due to
the large deviation $E-V$ of the energy from the barrier potential
$V$, due to which the $2\times 2$ Hamiltonian approximation inside
the barrier fails.

\vspace*{-0.5cm}
\section{superlattice}
\subsection{Dispersion relation}

\begin{figure}
     \begin{center}
         \includegraphics[height=4cm]{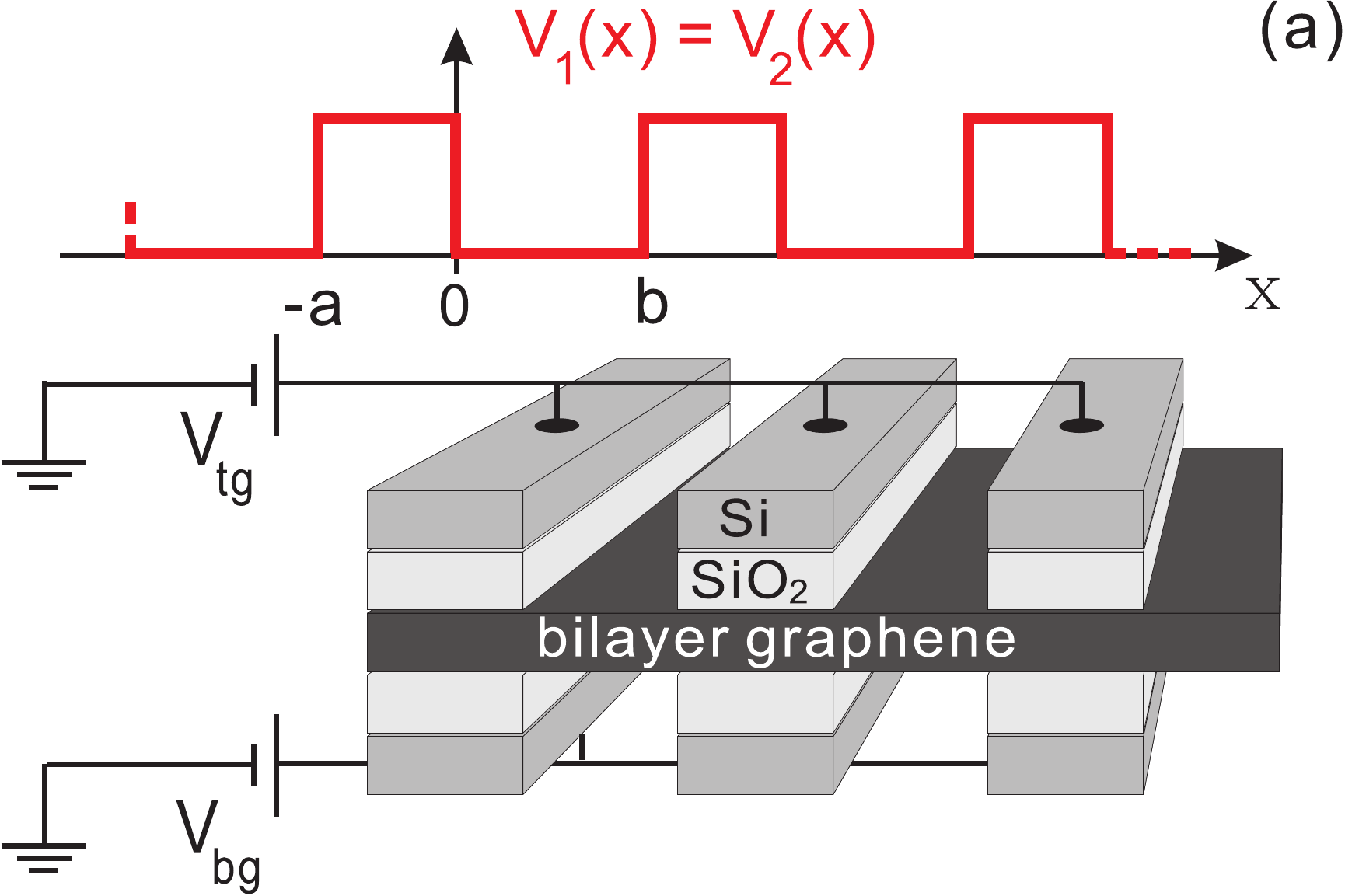}\\
         \includegraphics[height=4cm]{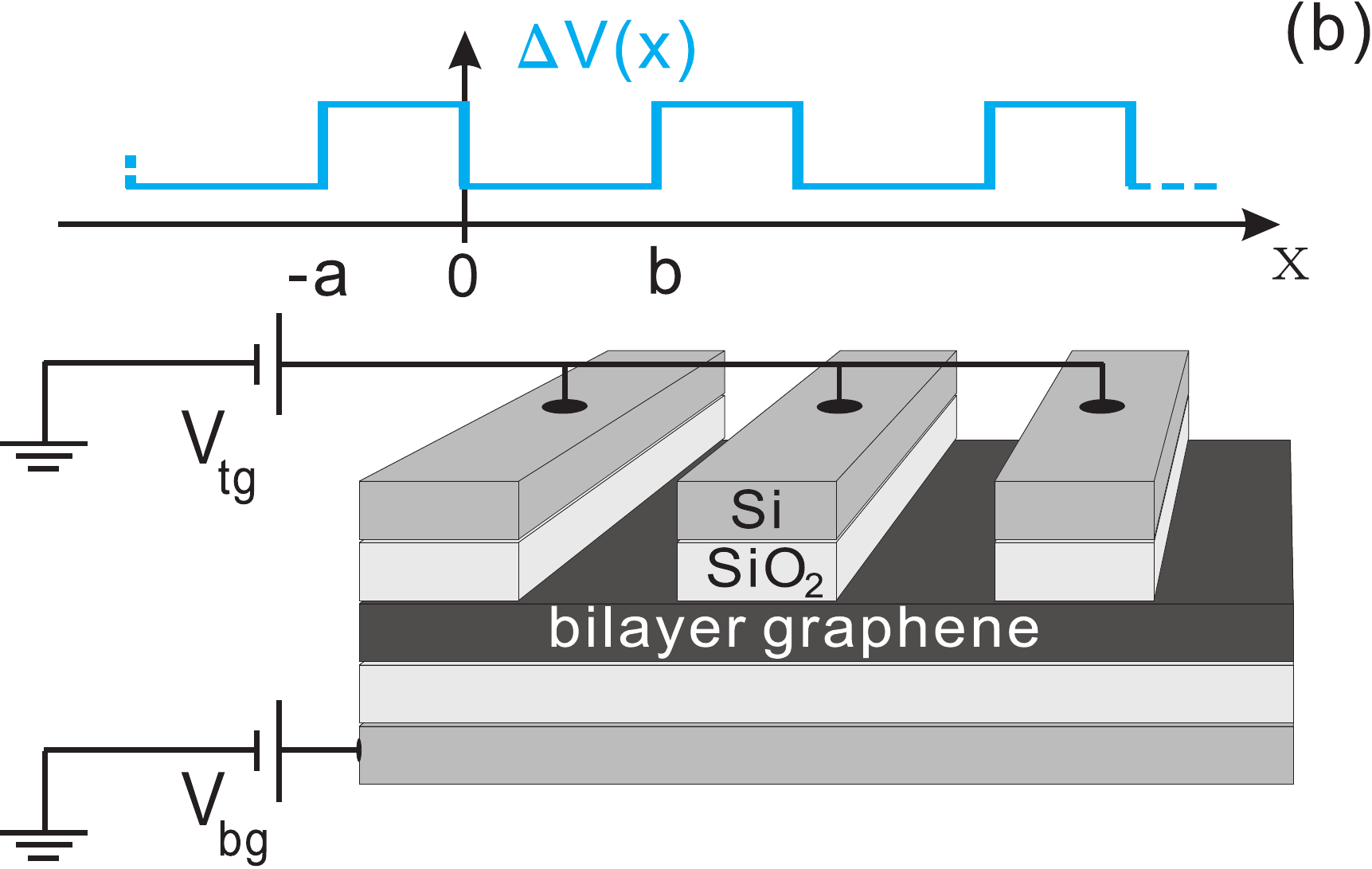}
    \end{center}
 \vspace*{-0.2cm}
    \caption{(Color online) Schematics of two experimental setups for realizing the three SL
    potentials we investigated. In panel {\bf (a)} the layer potentials $V_1$ and $V_2$
        are kept the same, the experimental setup shown
    can be used. The setup in panel {\bf (b)} can   establish a bias $\Delta = V_1-V_2$. In both 
    experimental setups the layer potentials 
    are controlled by the applied top   $V_{tg}$  and back  $V_{bg}$ gates.}
    \label{fig_pot1}
\end{figure}

The model we used for
 a superlattice (SL) in graphene is shown schematically in
Fig.~\ref{fig_pot1}. The 
electronic spectrum 
resulting from this periodic structure can be obtained by writing
the solution for the spinors as Bloch waves and applying the
continuity condition for the wave function at the potential steps.
In both barriers and wells 
the solutions are the ones for a constant potential and the
boundary conditions determine the matrix relation between the
wavefunction coefficients in the two regions. For a periodic
potential, Bloch's theorem applies with period $l = a + b$,
implying $\psi_k(x + l) = \psi_k(x) e^{i k l}$. Then referring to
Fig. \ref{fig_pot1} we 
obtain: $\psi(0-) = \psi(0+)$ and $\psi(-a) = \psi(b)e^{-i k l}$.
Writing the wave function in the regions of constant potential as a
matrix product $\Psi = \mathcal{G}\mathcal{M}\mathcal{A}$, 
labelling the coefficient matrices   inside the
barrier  regions as $\mathcal{A}_{1}$ and the ones  inside the well regions 
as $\mathcal{A}_{2}$, and applying the above boundary conditions
we obtain the matrix equations
\begin{equation}
     \mathcal{G}_1 \mathcal{A}_1 = \mathcal{G}_2 \mathcal{A}_2
\end{equation}
\begin{equation}
    \mathcal{G}_1 \mathcal{M}_1(-a) \mathcal{A}_1 = \mathcal{G}_2 \mathcal{M}_2(b) e^{- i k l} \mathcal{A}_2.
\end{equation}
Eliminating  $\mathcal{A}_1$ in Eqs. (10) and (11) leads to
\begin{equation}\label{eq_12}
     \left[ \mathcal{M}_1(-a) \mathcal{G}_1^{-1} \mathcal{G}_2
      - \mathcal{G}_1^{-1} \mathcal{G}_2 \mathcal{M}_2(b) e^{- i k l} \right] \mathcal{A}_2 = 0.
\end{equation}
Equating the determinant of Eq.~(\ref{eq_12}) to zero
\begin{equation}
     \det[\mathcal{M}_1(-a) \mathcal{G}_1^{-1} \mathcal{G}_2
      - \mathcal{G}_1^{-1} \mathcal{G}_2 \mathcal{M}_2(b) e^{- i k l}] =
      0.
\end{equation}
The solution of Eq.~(14) gives the energy spectrum or dispersion
relation.
From this determinant we search for the zeros of Eq.~(14) using the
Newton method and obtain the dispersion relation.
%
\begin{figure*}[ht]
    \begin{center}
				\subfigure{\includegraphics[height=5cm,width=6cm]{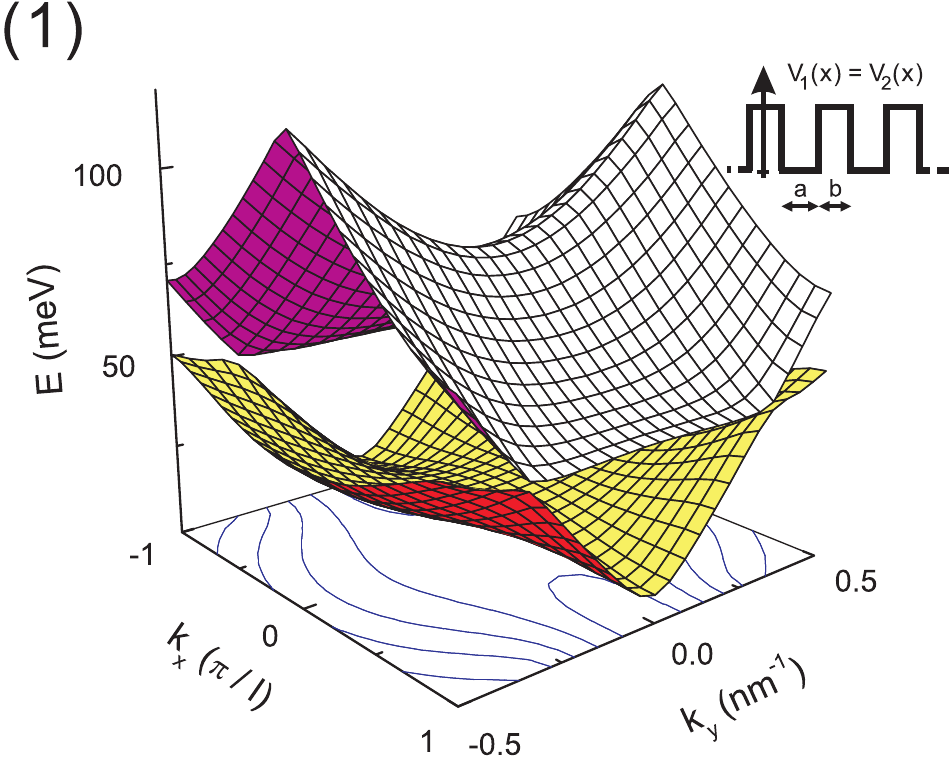}}
        \subfigure{\includegraphics[height=5cm,width=5cm]{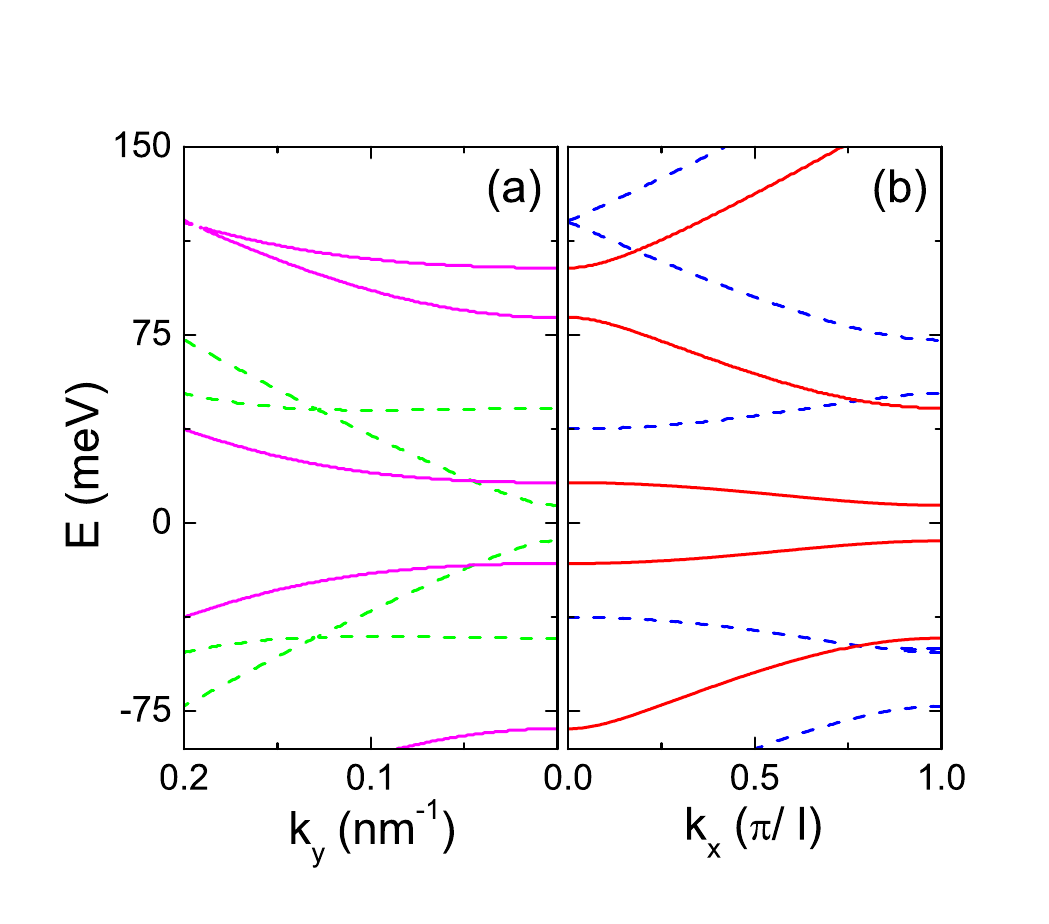}}
        \subfigure{\includegraphics[height=5cm,width=5cm]{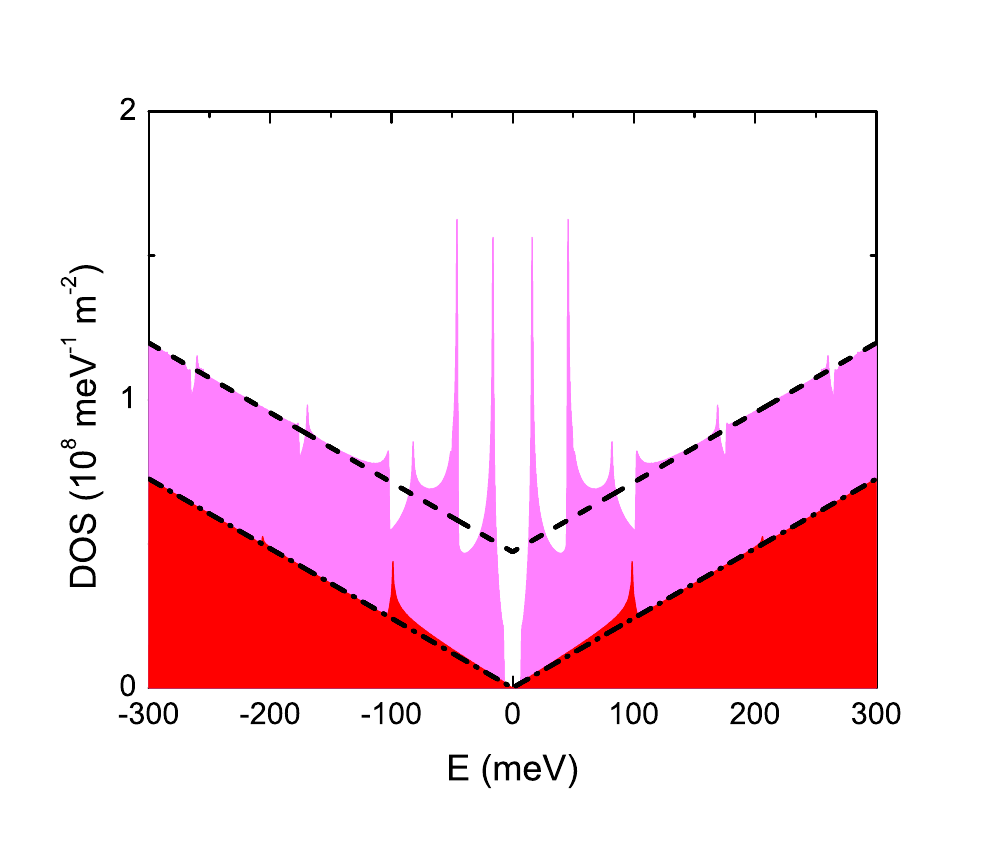}}\\
        \subfigure{\includegraphics[height=5cm,width=6cm]{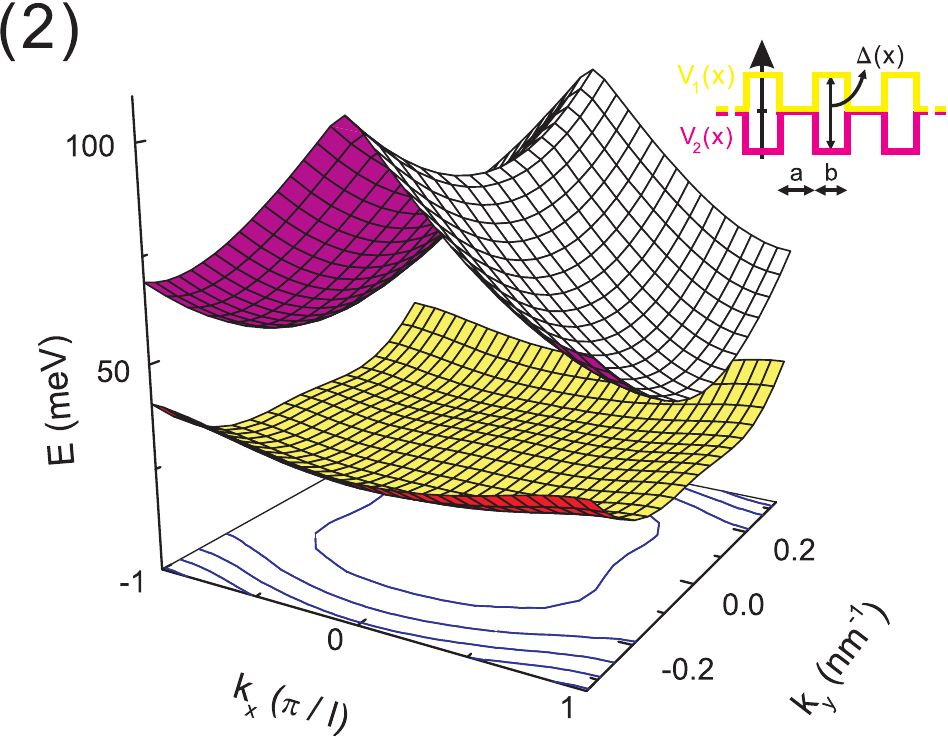}}
        \subfigure{\includegraphics[height=5cm,width=5cm]{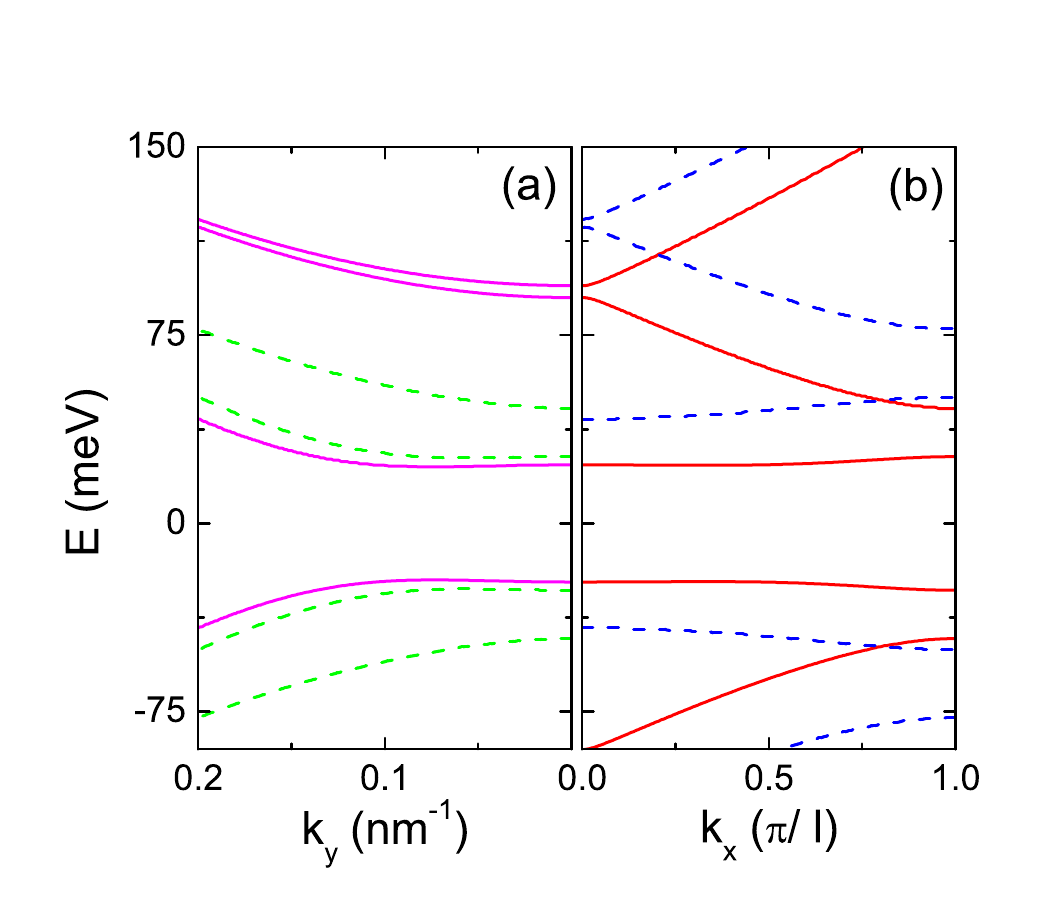}}
        \subfigure{\includegraphics[height=5cm,width=5cm]{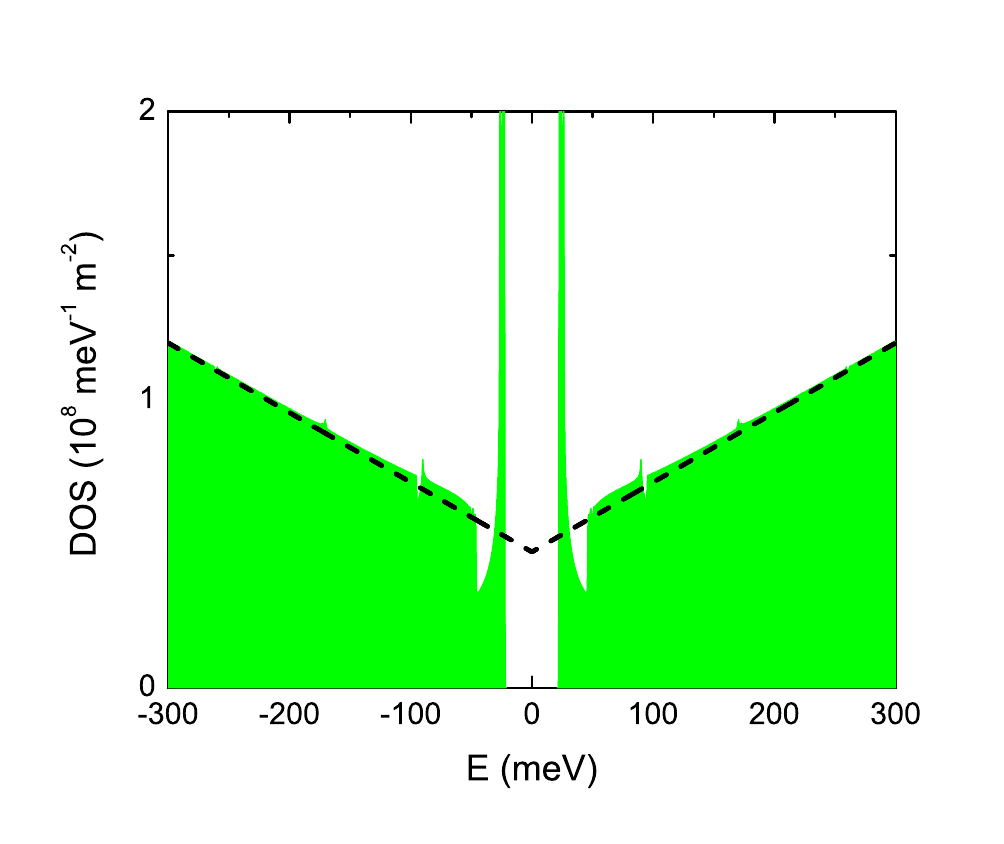}}\\
        \subfigure{\includegraphics[height=5cm,width=6cm]{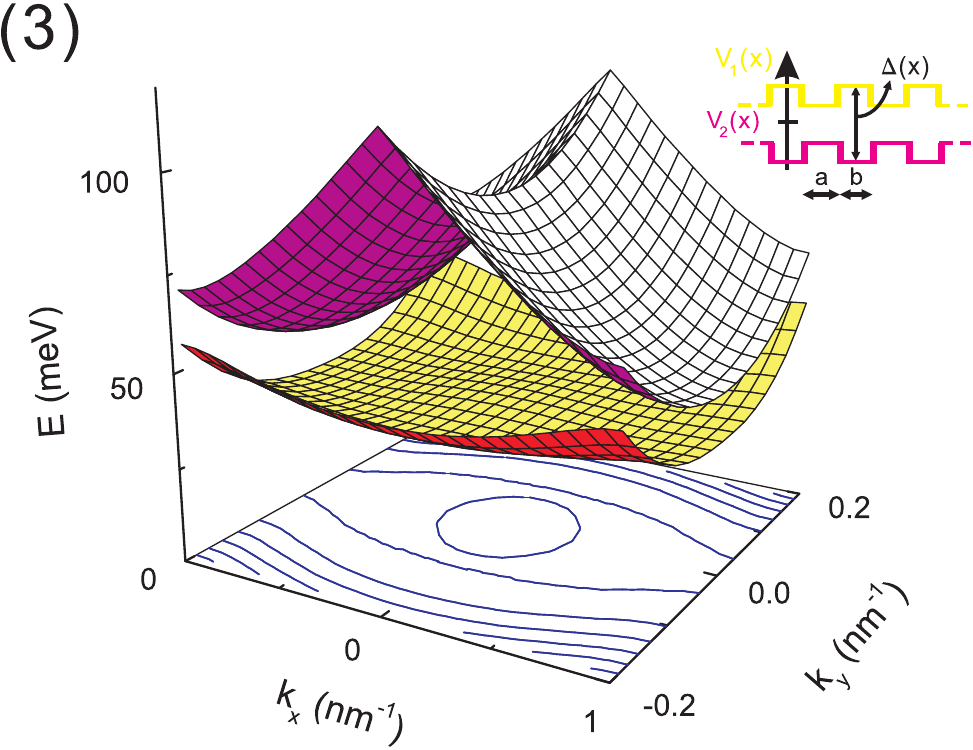}}
        \subfigure{\includegraphics[height=5cm,width=5cm]{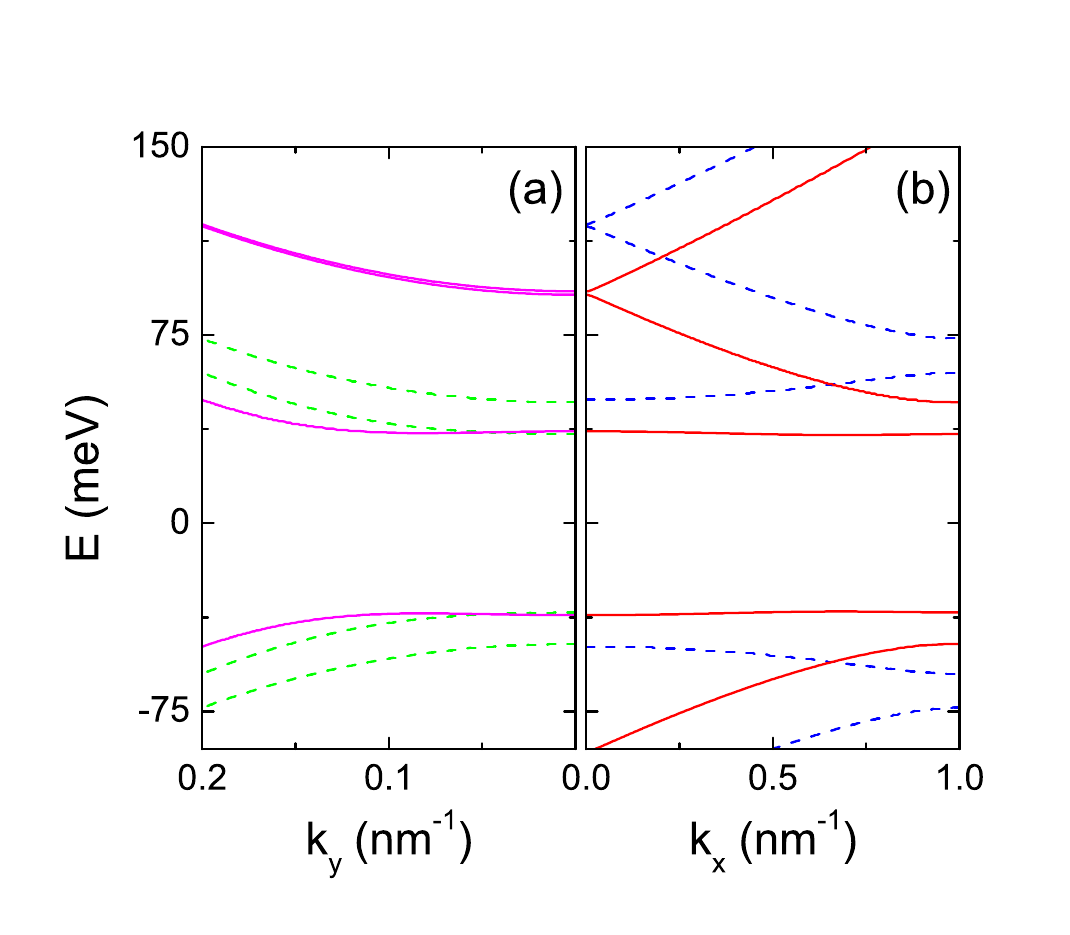}}
        \subfigure{\includegraphics[height=5cm,width=5cm]{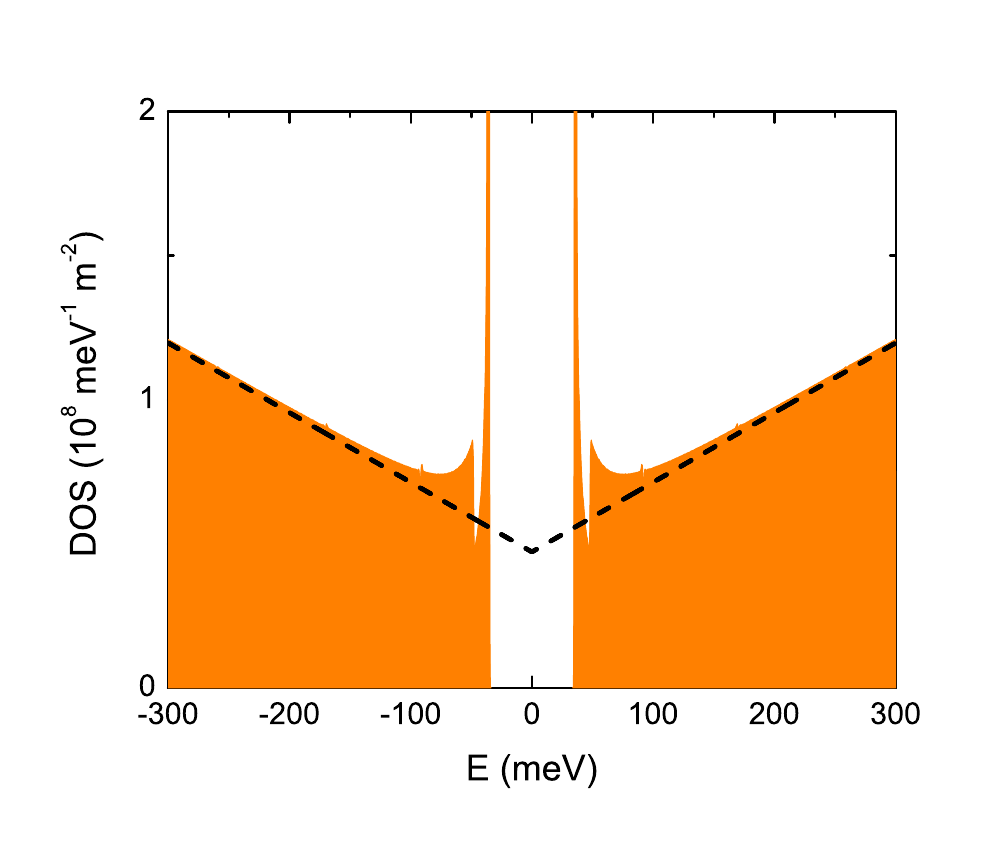}}\\
    \end{center}
\vspace*{-0.6cm}  \caption{(Color online) Dispersion relation and DOS for three 
types of SLs. 
{\bf (1)} The  barriers are $50$ meV high and the wells $-50$ meV
deep. {\bf (2)} The  barriers are biased by
$\Delta = 50$ meV,  the wells are unbiased, i.e., $\Delta = 0$ meV. {\bf (3)}  The  barriers are biased by $\Delta = 50$ meV and  the wells by $\Delta = 25$ meV. {\bf Left column:} energy vs $k_x$ and $k_y$ for $a = b = 10$ nm and $t_\perp = 390$ meV. 
Lines of constant energy, belonging to  the lower miniband, are projected onto the ($k_x, k_y$) plane. {\bf Middle column:} slices of the corresponding dispersion relation, (a) for constant $k_x = 0$ (solid magenta curves) and $k_x = \pi/l$ (dashed green curves), and (b) for constant $k_y = 0$ (solid red curves) and $k_y = 0.2$/nm (dashed blue curves). Only half the Brillouin zone is shown. {\bf Right column:} DOS for the corresponding 
SL. For the unbiased  SL {\bf (1)} we also show the DOS (red area)
for a SL with the same parameters 
on a single-layer graphene. 
The dashed (dash-dotted) curves show the bilayer (single-
layer) DOS in the absence of the SL potential.}
  \label{fig_disp3d1}
\end{figure*}

In Fig.~\ref{fig_disp3d1} we plot the dispersion relation versus
$k_x$ and $k_y$ for three different SLs. In the first one we take
the potential on the back and front gates to be the same $V = 50$
meV; between the strips the potential is $-50$ meV.
 In the other two we only vary the bias difference $\Delta$ between
the two layers:
 $\Delta = 50 $ meV in the barriers and $0$ meV  in the wells for the second SL and correspondingly
$\Delta = 50 $ meV and $25$ meV for  the third   one.
 The average
potential of both layers is kept constant. The parameters  used  are
$a = b = 10$ nm and the tunnel coupling\cite{ohta} $t_\perp = 390$ meV.
Only the first two minibands
 are shown in the left panels of Fig.~6. The middle  column shows cross sections of the dispersion relation 
for constant  $k_x$ in panels (a) and
constant $k_y$ in panels (b).

Applying the first type of SL potential shows the formation of
subbands.
For this first type one can easily find an analytical expression for the one-dimensional case, i.e. $k_y = 0$, the formula for the dispersion calculated from the four-band Hamiltonian is (union of formula with $+$ and $-$)
\begin{equation}
	\cos(k L) = \cos(\alpha_{\pm,1} a)\cos(\alpha_{\pm,2} b) - G_\pm \sin(\alpha_{\pm,1} a)\sin(\alpha_{\pm,2} b)
\end{equation}
Where $\ve_j = E - V_j$ in region $j$
\begin{equation}
	G_\pm = \frac{(\alpha_{\pm,2}^2\ve_1^2 + \alpha_{\pm,1}^2\ve_2^2)}{2\alpha_{\pm,1}\alpha_{\pm,2}\ve_1\ve_2}
\end{equation}
The analog of this formula for the two-band approximation of the Hamiltonian is the same formula with $\alpha_{\pm,j} = \sqrt{\pm \ve_j}$ instead of $\alpha_{\pm,j} = \sqrt{\ve^2 \pm \ve_j t_\perp}$

In contrast with the gapless spectrum of SLs on  single-layer
graphene, 
here a bandgap is found for $k_y = 0$. 
This is in agreement with the fact that for the transmission
through a barrier there is no perfect transmission for
perpendicular incidence while in single-layer graphene there is. 
The second SL potential has the same barrier and well parameters
as Fig. 2(b), the resonance at $k_x \simeq 0.16$ nm$^{-1}$ meV which we saw in
this double barrier system seems to correspond to the energy value
of the first band, $E \simeq 25$ meV. Also, for the lowest band the mexican-hat
energy profile of biased bilayer graphene 
is retained in the $k_y$ direction. 
In the third SL potential the gaps between the subbands are
smaller than those  of the second SL and the dispersion relation
resembles more the (folded) one of bilayer graphene without any SL potential 
but with an applied constant potential difference. 
The DOS of the latter two SLs shows large van Hove peaks at  energies corresponding to the lowest band, there also the velocity is zero and localized states form.
 \vspace*{-0.5cm}
\subsection{Density of states}
 To understand part of the behavior of  carriers in a SL we evaluate the density
 of states (DOS) $D(E)$. In the reduced-zone scheme it is given by
\begin{equation}
 \hspace*{-0.2cm}
    D(E)  = \frac{4A}{\pi^2} \sum_n \int_{0}^{\pi/l} d k_x \int_{0}^{\infty}
    d k_y \delta(E - E_n(k_x,k_y)),
\end{equation}
where  $A$ is the surface area. The  integral is evaluated numerically by
converting it to a sum in the manner
\begin{equation}
    \int_{0}^{\pi/l} \ud k_x \int_{0}^{\infty} \ud k_y \approx \big(\frac{\pi}{N_x l}\big)
    \big( \frac{k_{max}}{N_y}  \big) \sum_{k_x = 0}^{\pi/l} \sum_{k_y = 0}^{k_{max}},
\end{equation}
where  the $k_x$ and $k_y$ indices take the values
\begin{equation}
    k_x = \frac{n_x}{N_x}\frac{\pi}{l}, \,\, k_y = \frac{n_y}{N_y}k_{max}, \,\, n_x(n_y) = 1 \cdots N_x(N_y).
\end{equation}
The cutoff $k_{max}$ for $k_y$  is chosen sufficiently large,
we took $k_{max} = 2$ nm$^{-1}$. In
addition, we  replace the $\delta$ function in Eq.~(15) by a
gaussian,
\begin{equation}
    \delta(E-E_n(k_x, k_y)) \approx (1/\sqrt{2\pi}) 
    \ e^{ - [E-E_n(k_x, k_y)]^2/2\sigma^2},\\
\end{equation}
and choose $\sigma$ small but sufficiently large to compensate for
the discretization of $k_x$ and $k_y$, i.~e.~we took $\sigma = 0.03$ meV.
The evaluated DOS is shown in the right panel of Fig.~6. In these
 figures the magenta, green and orange areas are for bilayer SLs and the red one for a single-layer SL. The dashed and dash-dotted curves show the DOS for
single-layer ($D_s$) and bilayer ($D_b$) graphene in the absence of the SL
potential given by
\begin{equation}
\begin{aligned}
    D_{s}(E) & = |E| /h v_F , \\
    D_{b}(E) & =(|E| + t_\perp/2) /hv_F,
\end{aligned}
\end{equation}
where we used the usual tight-binding Hamiltonian\cite{louie} for
single-layer graphene  and the one given by Eq.~(\ref{eq_1}) for
bilayer graphene. The peaks in the DOS have the typical $1/\sqrt{E-E_0}$ behavior of 1D subbands.
\section{Summary and concluding remarks}

We evaluated the electronic transmission and conductance through a
finite number of bilayer graphene barriers.  Further, we obtained 
the dispersion relation and the DOS for a periodically biased
bilayer, i.e., a  bilayer in the presence of a SL potential. With
the rapid progress in the field we expect that such a periodic
biasing will soon be realized experimentally. 
Since the elastic
mean free path of carriers in high-mobility graphene layers can be
of the order of hundreds of nanometers, a ballistic behavior can
be expected to be observable on the length scale of the periodic
structures discussed here.

For some transmission and conductance results 
we used both the 
four-band  Hamiltonian given by Eq.~(\ref{eq_1}) as well as the
reduced two-band Hamiltonian given by Eq.~(3), cf. Fig. 4 and 5. 
We consider the former results as more accurate than the latter
ones, since the graphene bilayer spectrum obtained from the
four-band Hamiltonian is known to give a better agreement with
both experimental data and theoretical tight-binding calculations\cite{ohta}.

For zero bias the dispersion relation shows a finite gap for
carriers with zero momentum in the direction parallel to the
barriers in contrast to the well-known results\cite{cast,louie} for
single-layer graphene, cf. Fig. 6. A gap also appears for a finite
bias, cf. Fig. 6. We also contrasted the DOS 
for bilayer graphene with the corresponding one for
single-layer graphene, cf. Fig. 6. We expect that all these
results will  be tested experimentally in the near future.

\acknowledgments This work was supported by IMEC, 
the
Flemish Science Foundation (FWO-Vl), the Belgian Science Policy
(IAP), the Brazilian Council for Research (CNPq), and the Canadian
NSERC Grant No. OGP0121756.
\section{Appendix}
We assume solutions of the form $\Psi_C(x, y) = \phi_C(x)e^{ik_y y}, C=A,B$. Then 
Eq.~(1) and  Schr\"odinger's equation $\mathcal{H} \psi = E \psi$  lead to the following equations
\begin{subequations}\label{eq_19}
\begin{align}
    - i (\pdje{x} - k_y) \phi_B & = (\ve'-\delta) \phi_A - t' \phi_{B'}, \label{eq_1a}\\
    - i (\pdje{x} + k_y) \phi_A & = (\ve'-\delta) \phi_B, \label{eq_1b}\\
    - i (\pdje{x} + k_y) \phi_{A'} & = (\ve'+\delta) \phi_{B'} - t' \phi_A, \label{eq_1c}\\
    - i (\pdje{x} - k_y) \phi_{B'} & = (\ve'+\delta) \phi_{A'}, \label{eq_1d}
\end{align}
\end{subequations}
where $\ve' \mp \delta=\ve - (u_0 \pm \delta), u_{1} = u_0 + \delta$, and
$u_{2} = u_0 - \delta$.
We solve Eq.~(\ref{eq_1b}) for $\phi_B$ and Eq.~(\ref{eq_1d}) for $\phi_{A'}$
and substitute the results in Eqs.~(\ref{eq_1a}) and (\ref{eq_1c}). This gives
\begin{subequations}
\begin{align}
    (\pdje{x}^2 - k_y^2) \phi_A & = -(\ve'-\delta)^2 \phi_A + t'(\ve'-\delta)\phi_{B'}, \\
    (\pdje{x}^2 - k_y^2) \phi_{B'} & = -(\ve'+\delta)^2 \phi_{B'} +
    t'(\ve'+\delta)\phi_{A}.
\end{align}
\end{subequations}
For the system of Eqs.~(\ref{eq_19}) and for constant potentials the spectrum
is determined by the equation
\begin{equation}\label{eq_23}
[-k^2 + (\ve'-\delta)^2][-k^2 + (\ve'+\delta)^2] - t'^2(\ve'^2 -
\delta^2) = 0.
\end{equation}
Solving it leads to  four 
bands ($ \epsilon_{kt'}^2=k^2 + \delta^2 + t'^2/2$)
\begin{equation}
    \ve_\pm^{'+} = \left[ \epsilon_{kt'}^2 
    \pm t' \sqrt{ 4k^2\delta^2/t'^2 + k^2 + t'^2/4 } \right]^{1/2}
\end{equation}
\begin{equation}
    \ve_\pm^{'-} = - \left[ \epsilon_{kt'}^2 
    \pm t' \sqrt{ 4k^2\delta^2/t'^2 + k^2 + t'^2/4 } \right]^{1/2}
\end{equation}
and four possible wave vectors $ \pm \alpha_\pm \simeq k_x = (k^2 -
k_y^2)^{1/2}$
\begin{equation}\label{eq_alpha1}
\hspace*{-0.3cm}\alpha_\text{\textcolor{blue}{$\pm$}} = \left[ \ve'^2
+ \delta^2 - {k_y}^2 \text{\textcolor{blue}{$\pm$}}
 \sqrt{ 4 \ve'^2 \delta^2 + t'^2 (\ve'^2 - \delta^2)} \right]^{1/2}.
\end{equation}
To obtain the general solution for the spinors we assume plane wave solutions
for $\phi_A = \phi^+_A + \phi^-_A$ of the form
\begin{equation}
    \phi^+_A = A e^{i \alpha_+ x} + B e^{-i \alpha_+ x}, \,
      \phi^-_A = C e^{i \alpha_- x} + D e^{-i \alpha_- x}.
\end{equation}
Then 
Eq. (\ref{eq_1b}) gives ($f^{
\text{\textcolor{blue}{$\pm$}}}_{\text{\textcolor{red}{$\pm$}}} =
[-i k_y \text{\textcolor{red}{$\pm$}}
\alpha_{\text{\textcolor{blue}{$\pm$}}}]/[\ve'-\delta]$.)
\begin{equation}
    \phi^\pm_B = f_+^\pm A e^{i \alpha_\pm x} + f_-^\pm B e^{- i \alpha_\pm
    x},
\end{equation}
with $A,B$ replaced by $C,D$, respectively, if the lower upper $-$ sign is used in $\phi$ and $f$.
Further,
Eq.~(\ref{eq_1a}) gives
\begin{equation}
    \phi_{B'} = h^\pm A e^{i \alpha_\pm x} + h^\pm B e^{-i \alpha_\pm
    x},
\end{equation}
\ \\
with $h^{ \text{\textcolor{blue}{$\pm$}}} = [(\ve'-\delta)^2 -
k_y^2 - {\alpha_{
\text{\textcolor{blue}{$\pm$}}}}^2]/[t'(\ve'-\delta)]$. Substituting
$\phi_{B'}$ in Eq.~(\ref{eq_1d}) gives
\begin{equation}
    \phi_{A'} = g_+^\pm h^\pm A e^{i \alpha_\pm x} + g_-^\pm h^\pm B e^{- i \alpha_\pm x}
\end{equation}
where $g^{ \text{\textcolor{blue}{$\pm$}}}_{\text{\textcolor{red}{$\pm$}}} =
 [i k_y \text{\textcolor{red}{$\pm$}}
 \alpha_{\text{\textcolor{blue}{$\pm$}}}]/[\ve'+\delta]$;
 the upper $\pm$ signs in $f$ and $g$ correspond to the subscripts of $\alpha$ and the lower ones to those 
 in front of $\alpha$. The eigenstates are
\begin{equation}\label{app_ev}
    \Psi^{ \text{\textcolor{blue}{$\pm$}}}_{ \text{\textcolor{red}{$\pm$}}} =
     N^{ \text{\textcolor{blue}{$\pm$}}}
     \kvecc{1}{f^{ \text{\textcolor{blue}{$\pm$}}}_{ \text{\textcolor{red}{$\pm$}}}}
     {h^{ \text{\textcolor{blue}{$\pm$}}}}
     {g^{ \text{\textcolor{blue}{$\pm$}}}_{ \text{\textcolor{red}{$\pm$}}} h^{ \text{\textcolor{blue}{$\pm$}}}}
     e^{\text{\textcolor{red}{$\pm$}} i \alpha_\text{\textcolor{blue}{$\pm$}} x + i k_y
     y}.
\end{equation}
$N^{ \text{\textcolor{blue}{$\pm$}}}$ is a normalization constant,
such that each state carries a unit current, and is given by
\begin{equation}
    {N^\pm}^2  = {t'(\ve^2 - \delta^2) \over
   2 W \alpha_\pm (t'(\ve'+\delta) + (\ve'-\delta)^2 - k_y^2 -
   \alpha_\pm^2)}.
\end{equation}
The solution $\Psi = (\Psi_A, \quad \Psi_B, \quad
\Psi_{B'}, \quad \Psi_{A'})^T$ can be rewritten
 in the matrix form
\begin{equation}\label{app_wavefct}
    \Psi = \kvecc{\Psi_A}{\Psi_B}{\Psi_{B'}}{\Psi_{A'}} =
    \mathcal{G} \mathcal{M}
    \kvecc{A}{B}{C}{D},
\end{equation}
with
\begin{equation}
    \mathcal{G} =
        \begin{pmatrix}
            1 & 1 & 1 & 1\\
            f^+_+ & f^+_- & f^-_+ & f^-_-\\
            h^+ & h^+ & h^- & h^-\\
            g^+_+ h^+ & g^+_- h^+ & g^-_+ h^- & g^-_- h^-\\
        \end{pmatrix}
\end{equation}
and
\begin{equation}
    \mathcal{M} =
        \begin{pmatrix}
            e^{i \alpha_+ x} & 0 & 0 & 0\\
            0 & e^{- i \alpha_+ x} & 0 & 0\\
            0 & 0 & e^{i \alpha_- x} & 0\\
            0 & 0 & 0 & e^{-i \alpha_- x}\\
        \end{pmatrix}.
\end{equation}
The columns of the matrix
product $\mathcal{G}\mathcal{M}$ are the
(unnormalized) eigenstates of our system.
\end{document}